\def\beg{\begin{equation}}
\def\eeq{\end{equation}}
\documentstyle[12pt]{article}
\begin{document}
\begin{center}
{\Large{\bf Fractional Charge in Quantum Hall effect}}
\vskip0.35cm
{\bf Keshav N. Shrivastava}
\vskip0.25cm
{\it School of Physics, University of Hyderabad,\\
Hyderabad  500046, India}
\end{center}

In 1976 Jackiw and Rebbi found 1/2 of a fermion number by using
Dirac equation. Schrieffer, in several proposals made an effort 
to suggest that there is a fractional charge. The calculations of Peierls distortion, Berry's phase and classical action were 
presented to accomodate the fractional charge in non-relativistic theory. Laughlin used the antisymmetry to define the charge 
density per unit area in a two dimensional system. In order to elliminate the area, Laughlin introduced the incompressibility 
which fixed the area and the odd number which determines the antisymmetry of the electron wave function, gave the charge. 
The antisymmetry relies on the odd number which can be equal to 
3 so that the charge became 1/3. We have used the orbital angular momentum and the spin of the electron to define the effective 
charge through the Bohr magneton which already has the charge 
of the electron. When both signs of the spin are used, 
$j={\it l} \pm s$, we generate a series of charges which are 
in agreement with those measured by Stormer in the quantum Hall 
effect experiments.
\vskip1.0cm
Corresponding author: keshav@mailaps.org\\
Fax: +91-40-2301 0145.Phone: 2301 0811.
\vskip1.0cm

\noindent {\bf 1.~ Introduction}

     In 1976, Jackiw and Rebbi[1], while considering the expansion
of the Fermi quantum fields in terms of eigen functions of the 
Dirac equation, introduced zero-energy solution in addition to 
positive and negative energy solutions, in 1+1 dimensions. They 
obtained conserved Fermi number current but the fermion number is 
found to be +1/2 for the soliton and -1/2 for the antisoliton. 
 Later, Su and Schrieffer[2] decided to introduce 
1/2 the charge of an electron in systems which undergo Peierl's distortion. Considerable effort was put to find the fractional 
charge in polyacetelene, $C-H_x$ chain. It was thought that $C-H_x$ chain will exhibit fractional charge but the ferromagnetic state 
of such a fractionally charged state is not known. In 1983, 
Laughlin wrote a wave function on the basis of antisymmetry from 
which it was suggested that the quasiparticles can have 1 or 1/3 
charge. Since antisymmetry requires that the exponent should be 
an odd number in order to generate a negative sign in the wave 
function with respect to interchange of any pair of particles. When 
this odd number is +1, interchanging a pair of particles changes the wave function $\psi$ to $-\psi$. For example, $z_1-z_2$ changes 
sign when 1 and 2 are interchanged. If the exponent is 3, we have $\psi \sim (z_1-z_2)^3$, which upon exchanging 1 and 2 also changes 
$\psi$ to $-\psi$. Laughlin defined the charge of the 
quasiparticles in such a way that the odd exponent 1 or 3 became 
the charge of the quasiparticles 1 or 1/3. Thus, a wave function 
was generated in which excitations will have a charge of 1/3. It 
was argued that this determines the wave function of quasiparticles, which cause steps in the quantum Hall effect experiments in semiconductors. Since the exponent must be an odd integer, the fractional charges produced are 1/3, 1/5, 1/7, etc. There was 
considerable belief that such charges are indeed found in the experimental data. If the charge became 1/3, what happened to the
remaining charge of 2/3 is not discussed. It is not clear whether 
there is a phase transition from e to e/3. In the Jackiw and Rebbi's theory the missing charge is carried by the soliton but in the Laughlin's theory one is concerned only with the wave function of 
the fractional charge.

While several authors were interested in seeing, if there is a spin-charge decoupling, we have found in 1986 that spin and charge are in fact coupled. This means that the knowledge of spin is needed to determine the value of the charge. The charge can be written in 
terms of spin and orbital angular momenta quantum numbers, ${\it l}$ 
and $s$,as,
\beg
{e_{eff}\over e} = {{\it l}+{1\over 2}\pm s\over 2{\it l}+1}
\eeq
It has its origin in the factors which multiply the Bohr magneton,
$\mu_B=e\hbar/2mc$ so that the charge $e$ can be replaced by 
$e_{eff}$. Usually, there is only one $g$ value associated with the Lande's formula which uses only one sign in L+S but here we are using half the value with both signs, $j={\it l}\pm s$. When the charge $e_{eff}=0$,
we see that there is a particle with spin $\pm 1/2$. The zero charge also corresponds to zero energy, so that at this point there is a charge-density wave. Thus we predict a particle which is electrically
neutral but has a spin. Not only that, but there is an effort to identify these objects in the experimental data of quantum Hall effect and it is claimed that such quasiparticles really exist.

In this paper, we describe the suggestions of Schrieffer about the fractional charges, present a brief review of Laughlin's wave function and our own work on the fractional charges.

\noindent{\bf 2.~~Schrieffer's suggestions}

{\it (a) Peierls distortion.}

     It has been suggested by Su and Schrieffer[2] that in the case
of dimerization, if the electrons were spinless, the charge of a kink will be $+{1\over 2}|e|$. In the case of spin, charged kinks are spinless and neutral kinks have spin $1/2$ in contrast to $\pm e$
charge for electrons and holes in semiconductors and spin $1/2$, i.e.,
\beg
charge = \pm {1\over 2}e, spin =0;
\eeq
and
\beg
charge =0, spin =1/2;
\eeq
and the usual,
\beg
charge = \pm e, spin=1/2;
\eeq
excitations are possible.  In the one-third filled band case the 
charge deficit is $(1/3)e$ or $(2/3)e$ and one is left with a 
fractional charge,
\beg
Q = \pm {1\over 3} |e|\,\,\,\,\, or\,\,\, {2\over 3}|e|,
\eeq
but there is no way of knowing, what determines whether there is dimerization or trimerization. For commensurability $n$, the kinks 
have charge which is an integer multiple of $\pm e/n$.

     Consider three types of distortions with sites numbered from 
-1, 0, 1, 2, ...,8. The dimerizations occur as (-1,0), (2,3), (5,6) which means that atoms on sites -1 and 0 are paired, those on 2 
and 3 are paired and (5,6) are paired. This is called A dimerization. Similarly (0,1), (3,4), (6,7) is called B and (1,2), (4,5), (7,8) is called C. There are two types of kinks, type I is A to B, B to C or 
C to A and type II are A to C, C to B or B to A. The primitive charge 
of type-I kink  is $+(2/3)|e|$ and that of type-II kink is $-(2/3)|e|$. 
In the case of type-I kinks, there is a localized state, $\varphi_l$ 
and symmetrically a state $\varphi_u$ in the lower half of the upper gap. If $\varphi_l$ has a charge of $(2/3)|e|$ with all spins paired, 
the kink spin is zero. If $\varphi_l$ is slightly occupied, the kink 
has charge $-(1/3)|e|$ and spin 1/2. If $\varphi_l$ is doubly occupied, the charge is $-(4/3)|e|$ and the spin is zero. For type-II kinks, the localized states $\varphi_l$ and $\varphi_u$ are below and above the 
gap centers, respectively. If $\varphi_l$ is unoccupied, the kink has charge $+(4/3)|e|$ and spin 0. For $\varphi_l$, singly occupied, the charge is $(1/3)|e|$ and spin is (1/2), while for $\varphi_l$ doubly occupied, the charge is $-(2/3)|e|$ and spin is zero. In other words, fractional charges, with or without spin occur.

     In the above discussion, we have not considered the spin=0, charge=0; spin=0, charge $\pm$ 1 or spin = 1/2, charge = $\pm$ 1/2. 
The spin = zero,  charge =0 is perfectly correct because it is a neutral boson; spin=0, charge =$\pm$ 1 is a charged particle without spin 
which is also very reasonable and spin =1/2, charge = $\pm$ 1/2 has unusual symmetry. However, due to Goldstone theorem, only spin=0 quasiparticle should emerge in a Peierls distortion. Therefore, the fractional charge which may not be a boson violates the Goldstone theorem.

{\it(b) Berry's phase.}

     Arovas et al[3] calculate the change of phase $\gamma$ of a state having a quasihole localized at $z_o$. As $z_o$ adiabatically moves around a circle of radius R enclosing flux $\phi$, the 
accumulated phase factor is, 
\beg
{e*\over \hbar c}\oint {\bf A}.d{\bf l}= 2 \pi ({e*\over e})({\phi
 \over \phi_o}).
\eeq
This is the phase gained by the quasiparticle of charge e* in going around a loop. It may be noted that whether e* is changed or {\bf A}
is changed, will be indistinguishable since only the product 
e*{\bf A} occurs in the phase factor. It will be desirable to 
obtain a link with the wave functions suggested by Laughlin[4] for 
the quasiparticles of fractional charge. Laughlin's calculation is concerned with the charge per unit area and not only the 
quasiparticle charge so that e* should be replaced by e*/m$a_o^2$, 
where m is an odd integer and $a_o^2$=$\hbar c/eB$, which involves 
the magnetic induction, B. The charge $e$ in absence of a field is linked to the Bohr radius, $a_B$. The corrected form of the phase 
factor must have $ma_o^2$. Since, the phase must be dimensionless, 
the corrected form of the phase should be,
\beg
\gamma = 2\pi ({e*\over e})({a_B^2\over ma_o^2})({\phi\over \phi_o}).
\eeq
This means that the phase factor is linearly dependent on the 
magnetic field. At the field of a few Tesla, $a_o\sim 10^{-4} \,\, cm$,while $a_B\sim 10^{-8}\,\, cm$ so that $(a_B/a_o)^2\sim 10^{-8}$.
Therefore, the phase factor $\gamma$ is a negligibly small number. If $\omega$ is the frequency and $\pi/a_o$ is the wave vector, then the velocity is $\omega a_o/\pi$  and $\gamma$ is not independent of the velocity. The charge of the quasiparticle is $\pm e*=e/m$ so that it 
is possible to obtain a fractional charge,
\beg
Q*=\pm \nu e.
\eeq
as a possibility.

{\it(c) Classical action.}

     Kivelson et al[6] have calculated the classical action and found that when the real part is less than a critical value, there is a phase transition. The fractional values of the charge such as $e/3$ satisfy the condition of a phase transition so that a phase transition is possible for a fractional charge. Here also we have found[7] that whether $e$ is being changed to $e*$ or ${\it l}_o$ is being changed to ${\it l}_o/\nu^2$, are indistinguishable.

\noindent{\bf3. ~~ Laughlin's theory}

     Laughlin utilized the antisymmetry property of the wave function
of fermions to define a fractional charge. Let us consider a wave function of the form,
\beg
\psi = (z_1-z_2)^mf_s
\eeq
where $f_s$ is a function which is symmetric with respect to 
interchange of 1 and 2, i.e., there is no change in $f_s$ when the particles 1 and 2 are interchanged, $f_s$(1,2)=$f_s$(2,1). This property is not associated with the anticommutators for fermions. On the other hand, when m=1, the factor $z_1-z_2$ changes sign when we interchange 1 and 2. Therefore, the factor $z_1-z_2$ is associated with the antisymmetry. Usually, it is enough to have m=1 but it is clear that the antisymmetry is obtained as long as $m$ is an odd integer.
For example, for $m$=3 also the factor containing $z_1-z_2$ will change sign with respect to interchange of 1 and 2. The factor $(z_1-z_2)^m$ is therefore associated with the 
antisymmetry of the wave function with respect to interchange of 
1 and 2 as long as $m$ is an odd integer.
Laughlin has found that the fluid density is,
\beg
\rho_m={1\over m2\pi a_o^2}
\eeq
so that the charge density is $e/[m2\pi a_o^2]$. Laughlin 
interpreted that the charge has become $e/m$ or e/1, e/3, e/5, 
etc. For this interpretation, it is possible to explain the 
experimental data on noise measurements as well as plateaus in 
the quantum Hall effect. An equally valid interpretation is that 
charge does not fractionalize but $a_o^2$ does. That means, 
instead of $e/m$ and $2\pi a_o^2$, we select $e$ and 
$2\pi (ma_o^2)$ in the above expression for density. The 
antisymmetry property or the exactness of Laughlin's calculation 
remain unaffected by this change in the interpretation. The square 
of the magnetic length is given by,
\beg
a_o^2={\hbar c\over eH_o}.
\eeq
Therefore, $ma_o^2$ has the magnetic field as the multiplier. 
Therefore, the quantity which appears in the charge density becomes $ma_o^2H_o$. Therefore, $m$ need not be attached to the charge. 
The energy of the electrons in a magnetic field is given by 
$g\mu_BH.S$ where charge $e$ is included in $\mu_B$. Therefore, 
four distinct quantities are involved. If any one of these is 
changed, the blame can easily go to any of the other three.

\noindent{\bf4.~~Theory}.

   We[9-15] use a vector model of ${\bf l}$ and ${\bf s}$ to 
define ${\bf j}$. In the Lande's formula only ${\bf J= L + S}$
is used so that there is only one value but we use 
${\bf j = l \pm s}$ so that there are two values,$g_{\pm}$ which is different from Lande's. When we define the Bohr magneton, the effective charge of the electron becomes,
\beg
e_{eff}= {{\it l}+({1\over 2})\pm s\over 2{\it l} + 1}e
\eeq
For various values of ${\it l}$ = 0, 1, 2, 3, ... $\infty$, this expression gives two series of charges, one for + sign and the other for minus sign. This predicted series is exactly the same as that measured by Stormer[6] in the quantum Hall effect. The magnetic moment of the quasiparticles is given by,
\beg
\mu_{eff}= {e_{eff}\hbar\over 2mc}.
\eeq
Therefore, when we go from the classical Hall effect to the quantum Hall effect, the charge of the quasiparticles changes from $e$ to $e_{eff}$. 
A large amount of data on the quantum Hall effect has been examined by us and our theory is found to be correct in all the cases.

\noindent{\bf5.~~ Conclusions}.

     In the case of an incompressible system $a_o$ has been left out so that the effect of $m$ is to change the charge to $e/m$ where $m$ is an odd integer associated with the antisymmetry of the wave function. On the other hand, we can keep $e$ unchanged and throw the effect of $m$ on $a_o^2$ so that the product $e/ma_o^2$ is unchanged. Since $a_o$ depends on the magnetic field, $e/(ma_o^2)=e^2H_o/(m\hbar c)$, changing $e$ is equivalent to changing $H_o$. Hence, the change in $H_o$ depends on ${\it l}$ and $s$ and we obtain spin dependent flux quantization. We also obtain a quasiparticle which has zero charge and spin =1/2.

\vskip1.25cm

\noindent{\bf6.~~References}
\begin{enumerate}
\item R. Jackiw and C. Rebbi,Phys. Rev. D{\bf13}, 3398(1976).
\item W. P. Su and J. R. Schrieffer, Phys. Rev. Lett. {\bf 46},738(1981).
\item D. Arovas, J. R. Schrieffer and F. Wilczek, Phys. Rev. Lett. {\bf 53}, 722 (1984).
\item R. B. Laughlin, Phys. Rev. Lett. {\bf 50}, 1395 (1983).
\item K. N. Shrivastava, cond-mat/0211351.
\item S. Kivelson, C. Kallin, D. P. Arovas and J. R. Schrieffer, Phys. Rev. Lett. {\bf 56}, 873 (1986).
\item K. N. Shrivastava, cond-mat/0211621.
\item K. N. Shrivastava, cond-mat/0210238.
\item K. N. Shrivastava, Natl. Acad. Sci. Lett. (India) {\bf 9},145(1986).
\item K. N. Shrivastava, Phys. Lett. A{\bf113}, 435 (1986);A{\bf 115},459(1986)(E).
\item K. N. Shrivastava, Mod. Phys. Lett. B {\bf 13},1087(1999).
\item K. N. Shrivastava, Mod. Phys. Lett. B {\bf 14}, 1009(2000).
\item K. N. Shrivastava, p.235 in, Frontiers of Fundamental Physics 4, edited by B. G. Sidharth and M. V. Altaisky, Kluwer Academic, New York 2001.
\item K. N. Shrivastava, Superconductivity: Elementary Topics, World Scientific, New Jersey 2000.
\item K.N. Shrivastava, Introduction to quantum Hall effect,\\ 
      Nova Science Pub. Inc., N. Y. (2002).
\item H. L. Stormer, Rev. Mod. Phys. {\bf 71}, 875 (1999).
\end{enumerate}
\vskip0.1cm

Note: Ref.15 is available from:\\
 Nova Science Publishers, Inc.,\\
400 Oser Avenue, Suite 1600,\\
 Hauppauge, N. Y.. 11788-3619,\\
Tel.(631)-231-7269, Fax: (631)-231-8175,\\
 ISBN 1-59033-419-1 US$\$69$.\\
E-mail: novascience@Earthlink.net

\end{document}